\documentstyle[emulateapj]{article}

\slugcomment{To appear in PASP (2003 April).}

\begin{document}

\title{On the Progenitor of Supernova 2001du in NGC 1365\footnote{Based on
observations made with the NASA/ESA {\sl
Hubble Space Telescope}, obtained from the data archive of the Space Telescope
Science Institute, which is operated by the Association of Universities for
Research in Astronomy, Inc., under NASA contract NAS 5-26555.}}

\author{Schuyler D.~Van Dyk}
\affil{IPAC/Caltech, Mailcode 100-22, Pasadena CA  91125}
\authoremail{vandyk@ipac.caltech.edu}

\author{Weidong Li and Alexei V.~Filippenko}
\affil{Department of Astronomy, 601 Campbell Hall, University of
California, Berkeley, CA  94720-3411}
\authoremail{alex@astro.berkeley.edu, weidong@astro.berkeley.edu}

\begin{abstract}
{\sl Hubble Space Telescope\/} ({\sl HST}) WFPC2 archival images obtained years
prior to the explosion of the Type II-plateau supernova (SN) 2001du in NGC 1365
previously have been analyzed to isolate the progenitor star.  The SN site was
located using precise astrometry applied to the {\sl HST\/} images, with
significant uncertainty, leaving three possible progenitor candidates.  Images
of the fading SN have recently become publicly available in the {\sl HST\/}
archive, allowing us to pinpoint the SN's exact location on the pre-explosion
images.  We show that the SN occurred in very close proximity to one of the
blue candidate stars, but we argue that this star is not the actual progenitor.
Instead, the progenitor was not detected on the pre-SN images, and we constrain
the progenitor's mass to be less than $13^{+7}_{-4}\ M_{\odot}$.  This is
consistent with previous constraints on the progenitor masses of other Type
II-plateau supernovae (SNe), suggesting that such SNe arise from the iron core
collapse of massive stars at the lower extreme of the possible mass range.
\end{abstract}

\keywords{supernovae: general --- supernovae: individual (SN 2001du)
--- stars: massive --- stars: evolution --- stars: variables: other ---
galaxies: individual (NGC 1365)}

\section{Introduction}

Supernova 2001du was visually discovered by R.~Evans (2001) on August 24.7 (UT
dates are used throughout this paper) at the western end of the bar in the
spiral galaxy NGC 1365 (about 90\arcsec\ W and 10\arcsec\ S of the nucleus).
Smartt, Kilkenny, \& Meikle (2001) classified the SN as Type II; Wang et
al.~(2001), from a spectrum obtained on August 30 UT, confirmed the SN as Type
II-plateau (II-P) before optical maximum.  NGC 1365 has also been host to SN
1957C (of unknown type) and the Type Ic SN 1983V.  Using {\sl Hubble Space
Telescope\/} ({\sl HST}) observations of Cepheids, Silbermann et al.~(1999)
determined the distance modulus of NGC 1365 to be $\mu = 31.3{\pm}0.2$ mag,
corresponding to a distance of about 18.2 Mpc.

   Van Dyk, Li, \& Filippenko (2003) recently measured the absolute position of
SN 2001du and, applying an independent astrometric grid to pre-SN {\sl HST\/}
archival WFPC2 images, isolated three possible candidate progenitor stars
within the uncertainty of the measured SN position: one reddish star with a
detection in both the F555W and F814W bands, and two stars with only F555W
detections, implying bluer colors for these latter two stars.  Also, multi-band
WFPC2 images of the SN were obtained for {\sl HST\/} program GO-9041 (PI:
S.~Smartt) in 2001 November.  These images are now in the public domain, and in
this paper we analyze them and compare them to the pre-SN images to further
constrain the nature of the SN progenitor.  (Smartt et al.~2003 have also
recently analyzed their images and, in a somewhat similar manner, come to the
same basic conclusion that we reach here.)

\section{Analysis}

Pre-SN images, each of total exposure time 300 s in the bands F336W, F555W, and
F814W, were obtained by GTO-5222 on 1995 January 15.  Late-time multi-band
images of the SN were obtained by GO-9041 on 2001 November 26.  These include
deep images in F336W, F439W, F555W, F675W, and F814W, as well as shorter
exposures in F336W, F555W, and F814W.  The available {\sl HST\/} data are
summarized in Table 1, with total exposure times given for each band.

In Van Dyk et al.~(2003) we adopted the position of the SN, measured from an
image obtained with the YALO 1-m telescope at CTIO (and available on the 
Internet), as
$\alpha$ = 3$^h$33$^m$29${\fs}$15, $\delta$ =
$-$36{\arcdeg}08{\arcmin}31${\farcs}$5 (J2000), with total uncertainty
${\pm}0{\farcs}9$.  The SN site is located on the WF3 chip.  In Figure 1 we
reproduce Figure 21 from Van Dyk et al., showing the SN environment in the
F555W band.  In Figure 2 we show the late-time F555W image (the SN is nearly
centered on the PC chip for all the GO-9041 exposures).  As one can see, the
actual SN position is within the $0{\farcs}9$ error circle, which provides
additional confidence in the astrometric method used by Van Dyk et al. when
isolating the various SN sites\footnote{Applying the same astrometric method
used by Van Dyk et al.~(2003) to the mosaic of the deep F555W exposure, we
measure a position for SN 2001du to be $\alpha$ = 3$^h$33$^m$29${\fs}$14,
$\delta$ = $-$36{\arcdeg}08{\arcmin}31${\farcs}$3 (J2000), ${\pm}0{\farcs}2$
in each coordinate.}.

We initially used the routine HSTphot (Dolphin 2000a,b) to obtain photometry of
the objects imaged on the PC chip.  HSTphot automatically accounts for WFPC2
point-spread function (PSF) variations and charge-transfer effects across the
chips, zeropoints, aperture corrections, etc., and can return as output
magnitudes transformed from the {\sl HST\/} flight system to standard
Johnson-Cousins bands.  Van Dyk et al.~(2003) discuss the similarity in the
results from HSTphot, compared with those from the packages DoPHOT (Schechter,
Mateo, \& Saha 1993) and DAOPHOT/ALLSTAR (Stetson 1987, 1992).\footnote{In
fact, here we update the private communication by D.~C.~Leonard 2002, regarding
the very good agreement between the results of HSTphot photometry on images of
NGC 3351, compared with those obtained by Graham et al.~(1997) using DAOPHOT;
Leonard now finds ${\delta}V = 0.015{\pm}0.022$ and ${\delta}I =
0.036{\pm}0.018$ mag.}

The SN profile is saturated in the longer F555W ($V$), F675W ($R$), and F814W
($I$) exposures.  Thus, no brightness measurement is possible.  The SN is not
saturated in the shorter F555W and F814W exposures, the long F336W ($U$) and
F439W ($B$) exposures, and also the short F336W exposure.  However, the results
from HSTphot for the F336W exposures varied, depending on how these exposures
were combined with the other bands as input into the routine, and the results
from the long and short exposures for this band were also discrepant with each
other (the results ranged from about $m_{\rm F336W}=18.9$ to 18.7 mag for
different runs of HSTphot).  We do not know why this occurred, but we speculate
that it possibly may be due to the low signal-to-noise ratio for both exposure
sets.  For F336W we then applied to both the long and short exposures
DAOPHOT/ALLSTAR within IRAF\footnote{IRAF (Image Reduction and Analysis
Facility) is distributed by the National Optical Astronomy Observatories, which
are operated by the Association of Universities for Research in Astronomy,
Inc., under cooperative agreement with the National Science Foundation.} as a
check, using a TinyTim PSF (Krist 1995), the zeropoint from Holtzman et
al.~(1995), and aperture corrections ($\sim$0.00 mag) derived from artificial
star tests.  (Checks for the F439W and short F555W and F814W exposures with
DAOPHOT show good agreement with the HSTphot results.)

We obtained $m_{\rm F336W}=18.96{\pm}0.10$ and 18.87${\pm}0.04$ mag for the
short and long exposures, respectively.  We then averaged these DAOPHOT
magnitudes with the HSTphot magnitudes at F336W, and obtained $m_{\rm
F336W}=18.86$ mag.  The SN was quite red by late November 2001, with $m_{\rm
F336W} - m_{\rm F439W}=1.82$ mag.  Whether using the transformations from
flight system to Johnson-Cousins magnitudes from Holtzman et al.~(1995),
Dolphin (2000b), or via SYNPHOT applied to the Bruzual Spectral Synthetic Atlas
(see Van Dyk, Filippenko, \& Li 2002), this color is too red to obtain a
reliable transformation from $m_{\rm F336W}$ to $U$.  However, we do apply a
transformation, with uncertainty $\sim$0.2 mag.  The transformations for all
other bands are from Dolphin (2000b), applied internally within HSTphot.  The
final results for the photometry of SN 2001du in the {\sl HST\/} images are
presented in Table 2.

In Figure 3 we show preliminary, unpublished $BVI$ light curves for SN 2001du,
based on data from the YALO 1-m telescope (N.~Suntzeff \& K.~Krisciunas, 2003,
private communication), and we include the $UBVI$ photometry from the {\sl
HST\/} data.  For comparison we show the $B$ and $V$ light curves for the
plateau-type SN (II-P) 1999em in NGC 1637 (Leonard et al.~2002a).  It is
clearly evident that SN 2001du is indeed a SN II-P as well.  If this SN were a
linear-type event (II-L), it should have markedly declined in brightness by the
time of the {\sl HST\/} observations ($\sim$94 days since discovery).

\section{Discussion}

In Van Dyk et al.~(2003) we had isolated three faint objects, A--C, detected and
determined to be stellar by HSTphot, within the error circle around the SN
position.  Stars A and B, to the south and west, respectively, are both
blue, with $m_{\rm F555W}$ (${\sim}V$) $= 24.30 {\pm} 0.21$ mag and
$25.02 {\pm} 0.32$ mag, respectively, and no detections at
F814W (${\sim}I$).  (Blue, here, is relative, and includes stars with roughly
early-G spectral type or earlier, based on the detection limits; see below.)
Star C, to the east, is relatively red, with $V = 24.44
{\pm} 0.23$ and $V-I = 1.03{\pm}0.30$ mag.  We had assumed that, since
SN 2001du is of Type II-P, the most plausible of the three candidates is
the redder star.  However, that was argued to be still unlikely, based on the
assumed luminosity and color of Star C.

We have used the late-time SN images, specifically the $V$ image, and attempted
to match them up with the pre-SN images.  This was done by using the task {\it
geotran\/} in IRAF to transform both the late-time SN images to the pre-SN
images and vice versa.  We can therefore locate the position of the SN on the
pre-SN images as being 0.70$\pm$0.15 WF pixel ($0{\farcs}07$) northeast of the
position of Star B.  It would be tempting to assign Star B, then, as the
progenitor of SN 2001du.  However, (1) the offset of the SN position from Star
B (which corresponds spatially to 6.2 pc at the distance of NGC 1365) is
significantly large, and the accuracy of the measurement is appreciably high;
(2) SNe~II-P are thought to arise from the explosions of red supergiants; and
(3) no evidence exists that SN 2001du is peculiar in any way and that it arose
from a bluer, more compact star like the progenitor of SN 1987A (E.~Cappellaro,
2002, private communication).  Thus, it is more likely, as we previously had
suspected (Van Dyk et al.~2003), that the progenitor simply is not detected in
the pre-SN images.

We can use the detection limits of both the F555W and F814W pre-SN images to
constrain the nature of the progenitor.  We do this by inferring limits on the
luminosity and, therefore, the mass from the limits on the star's
apparent magnitude.  These HSTphot detection limits in the SN environment are
$m_{\rm F555W} > 25.19$ and $m_{\rm F814W} > 24.25$ mag.  Before converting
into absolute magnitude, we must first assess the amount of
extinction to the SN.  In Van Dyk et al.~(2003) we attempted to use stars in
the environment to estimate the reddening.  The Galactic extinction
toward NGC 1365 is $A_V=0.07$ mag (Schlegel, Finkbeiner, \& Davis 1998; and
NED\footnote{NED is the NASA/IPAC Extragalactic Database,
http://nedwww.ipac.caltech.edu.}).  We had found that stars in the larger SN
environment implied larger amounts of extinction, $A_V \approx 2.5$ mag.
(Unfortunately, although the post-SN images are deeper than the pre-SN ones,
the SN is far too bright, and its image is spread over too many surrounding
pixels, to perform a more thorough study of the environment.)

Using the light curve in Figure 3 we can actually derive more specific
information about the possible reddening toward the SN from the SN itself
(which is information that was not available to us previously).  The {\sl
HST\/} photometry of the SN on November 26 is about 94 days after discovery,
i.e., near the end of the light-curve plateau, where we can still estimate the
SN's color temperature with reasonable accuracy (e.g., Hamuy et al.~2001).
These colors are $B-V=1.42$ and $V-I=0.91$ mag.  For comparison, the colors of
the very well-studied Type II-P SN 1999em at the same age (about 95 days after
discovery) are $B-V=1.47$ and $V-I=0.89$ mag (Leonard et al.~2002a).  Leonard
et al.~and Hamuy et al.~(2001) conclude that SN 1999em is reddened by
$E(B-V)=0.10$ mag.  Since $B-V$ is affected by line blanketing in the $B$ band
(e.g., Hamuy et al.~2001), the small difference in $B-V$ (SN 1999em is 0.05 mag
redder than SN 2001du) could be due to differences in metallicity between the
two SNe, which might affect this color.  The marginally redder $V-I$ color of
SN 2001du (by 0.02 mag) may imply that this SN is slightly more reddened than
SN 1999em.  However, we consider these differences in color between the two SNe
to be small and likely within the uncertainties in the colors, so we adopt
$E(B-V) \approx 0.1$ mag for SN 2001du, as found for SN 1999em.

If we assume the Cardelli, Clayton, \& Mathis (1989) reddening law, then $A_V
\approx 0.3$ mag and $A_I \approx 0.2$ mag for SN 2001du.  For the distance
modulus $\mu=31.3$ mag, $M_V>-6.4$ and $M_I>-7.0$ mag for the progenitor star.
In a similar vein to the estimation of the progenitor mass limit for SN II-P
1999em by Smartt et al.~(2002) and SN II-P 1999gi by Smartt et al.~(2001), the
absolute magnitude limits can be converted to the luminosity of the likely
supergiant progenitor, assuming the full range of possible stellar surface
temperatures.  This is accomplished using a similar relation to equation (1) in
Smartt et al.~(2001, 2002), where we also have assumed the temperatures and
stellar bolometric corrections for supergiants from Drilling \& Landolt (2000)
and have computed, using SYNPHOT within IRAF and the Bruzual Spectral Atlas,
the conversions of $m_{\rm F555W}$ and $m_{\rm F814W}$ to Johnson $V$ for the
range in spectral type.  The luminosity limits for both F555W ($V$) and F814W
($I$) are shown in Figure 4.  Stars with luminosities brighter than these
limits (shown by the {\it heavy lines}) should have been detected in the pre-SN
images.

For comparison we have plotted (also in a similar fashion to Smartt et al.) the
model stellar evolutionary tracks for a range of masses from Lejeune \&
Schaerer (2001), assuming enhanced mass loss for the most massive stars and
that a metallicity $Z=0.04$ is appropriate for the SN environment.  Indication
for a metallicity somewhat greater than solar in this sector of NGC 1365 comes
from the abundance study by Roy \& Walsh (1997) of H~II regions in the galaxy.
The closest H~II regions to the SN site are their Nos.~20 and 21, where at this
radial distance from the nucleus, the O/H abundance is $\sim$9 dex or, in fact,
slightly greater, based on model assumptions (the solar O/H abundance is 8.8
dex; Grevesse \& Sauval 1998).

The luminosity limits at each band in Figure 4 show that the pre-SN images are
not sensitive to the bluest high-mass supergiants, so we cannot be absolutely
sure that such a star is not the progenitor.  If we exclude this possibility,
the $V$ band sets an upper limit to the progenitor mass of $\sim 13\
M_{\odot}$, although this ranges upward to as high as $\sim 20\ M_{\odot}$,
depending on the effective temperature.  The $I$ band is more sensitive to
less-massive red supergiants, ranging downward to $\sim 9\ M_{\odot}$ (possibly
even less) at the lowest effective temperatures.  These results from Figure 4
are roughly consistent with what we showed in Figure 22 of Van Dyk et
al.~(2003).  We therefore estimate that the upper mass limit for the SN 2001du
progenitor is $13^{+7}_{-4}\ M_{\odot}$.

This limit is consistent with the limit Leonard et al.~(2002b) find for SN II-P
1999gi ($15^{+5}_{-3}\ M_{\odot}$) and the $\sim 17\ M_{\odot}$ limit one finds
for SN II-P 1999em, if the distance to the SN 1999em host is assumed to be the
Cepheid distance, $11.5 \pm 0.9$ Mpc (Leonard et al. 2003).  Low masses for the
progenitors of SNe~II-P, relative to that indicated for SNe~II-L (linear)
and SNe~IIn (narrow), are also inferred from their optical properties
(e.g., Schlegel 1996; Filippenko 1997; Chugai \& Utrobin 2000) and suggested by
the inferred paucity of circumstellar matter from radio and X-ray observations
(e.g., Weiler et al.~1989; Schlegel 2001; Pooley et al.~2002).

\section{Conclusions}

Comparing {\sl HST\/} archival WFPC2 images obtained of the Type II-P SN 2001du
in NGC 1365 at late times with images obtained before explosion, we have
located the position of the SN progenitor star to within 1 WF chip pixel of the
candidate Star B identified by Van Dyk et al.~(2003).  This star, however, is
more than likely {\it not} the progenitor. We use the detection limits of the
pre-SN F555W and F814W images to place constraints on the progenitor's nature;
specifically, $M \lesssim 13^{+7}_{-4}\ M_{\odot}$.  (Smartt et al.~2003 find a
similar mass limit of $M = 15\ M_{\odot}$.)  This limit is consistent with what
has been found and inferred for other SNe~II-P.

The proximity of the SN to the blue Star B implies that the progenitor may have
been part of a small stellar association (the separation between the progenitor
and Star B is likely too large for the blue star to be a binary companion).
The faint emission underlying Stars A, B, and C identified by Van Dyk et
al.~(2003) suggests that an association complex may lie in the SN environment.
It would be interesting to image the SN again in multiple bands with {\sl
HST\/} at high spatial resolution, presumably with the Advanced Camera for
Surveys (ACS), when the SN has significantly dimmed.  The stars in the
environment of the fading SN would then become more obvious, and the age and
mass of the progenitor star might be further constrained, based on the
characteristics of its surviving neighbors.

\acknowledgements

This research made use of the NASA/IPAC Extragalactic Database (NED) which is
operated by the Jet Propulsion Laboratory, California Institute of Technology,
under contract with NASA, and the LEDA database (http://leda.univ-lyon.fr).
The work of A.V.F.'s group at UC Berkeley is supported by NASA grants AR-8754,
AR-9529, and GO-8602 from the Space Telescope Science Institute, which is
operated by AURA, Inc., under NASA contract NAS5-26555. We thank D.~C.~Leonard
and M.~Hamuy for useful discussions.  We also thank the referee, K.~Krisciunas,
for helpful comments and for providing unpublished ground-based photometry of
SN 2001du.


\begin{deluxetable}{lcclc}
\def\phmm{\phm{$-$}}
\tablenum{1}
\tablecaption{Summary of {\sl HST\/} Observations}
\tablehead{\colhead{Date} & \colhead{Filter}
& \colhead{Exp.~Time} & \colhead{{\sl HST}} & \colhead{SN} \nl
\colhead{(UT)} & \colhead{} & \colhead{(s)} & \colhead{Program}
& \colhead{Saturated?}}
\startdata
1995 Jan 15 & F336W & 300\tablenotemark{a} & GTO-5222 & \nodata \nl
            & F555W & 300 &          & \nodata \nl
            & F814W & 300 &          & \nodata \nl
2001 Nov 26 & F336W & 100 & GO-9041  & N       \nl
            &       & 460 &          & N       \nl
            & F439W & 400 &          & N       \nl
            & F555W &  40 &          & N       \nl
            &       & 600 &          & Y       \nl
            & F675W & 400 &          & Y       \nl
            & F814W &  40 &          & N       \nl
            &       & 600 &          & Y       \nl
\enddata
\tablenotetext{a}{The exposure times for all three bands in this 1995 Jan 15
dataset were erroneously listed in Van Dyk et al.~(2003) as 100~s. The
actual total exposure time of 300~s results from the combination of a trio 
of 100-s exposures. Note that HSTphot automatically reads the correct total 
exposure time from the images during processing, so the reported results are
unaffected by this error.}
\end{deluxetable}


\begin{deluxetable}{ccccc}
\tablenum{2}
\tablecolumns{5}
\tablewidth{17cm}
\tablecaption{Photometry of the SN~2001du {\sl HST\/} Images}
\tablehead{
\colhead{$U$} &
\colhead{$B$} &
\colhead{$V$} &
\colhead{$R$} &
\colhead{$I$} \nl
\colhead{[$m_{\rm F336W}$]} &
\colhead{[$m_{\rm F439W}$]} &
\colhead{[$m_{\rm F555W}$]} &
\colhead{[$m_{\rm F675W}$]} &
\colhead{[$m_{\rm F814W}$]} }
\startdata
18.6(0.2)\tablenotemark{a} & 16.72(0.06)  & 15.30(0.00) & \nodata & 14.39(0.00) 
\nl
[18.86]\tablenotemark{b} &   [17.04]      & [15.32]     & \nodata & [14.43] \nl
\enddata
\tablenotetext{a}{The value in parentheses for each entry in this row is the 
uncertainty in the transformed Johnson-Cousins magnitude.}
\tablenotetext{b}{The value in brackets for each entry in this row is the 
adopted flight-system magnitude, derived from HSTphot, DAOPHOT, or both.}
\end{deluxetable}


\begin{figure}
\figurenum{1}
\caption{The site of SN 2001du in the archival 300-s F555W image
from 1995 January 15.  The error circle has radius $0{\farcs}9$.  The three
stars, A--C, within the circle are indicated with tickmarks.  This is a
reproduction of Fig.~21 from Van Dyk et al.~(2003).}
\end{figure}


\begin{figure}
\figurenum{2}
\caption{SN 2001du, as seen in the 600-s F555W image from 2001 November 26,
shown to the same scale and orientation as in Figure 1.  The error circle
from Figure 1 is also shown.  The supernova is saturated in this image.}
\end{figure}

\clearpage

\begin{figure}
\figurenum{3}
\plotone{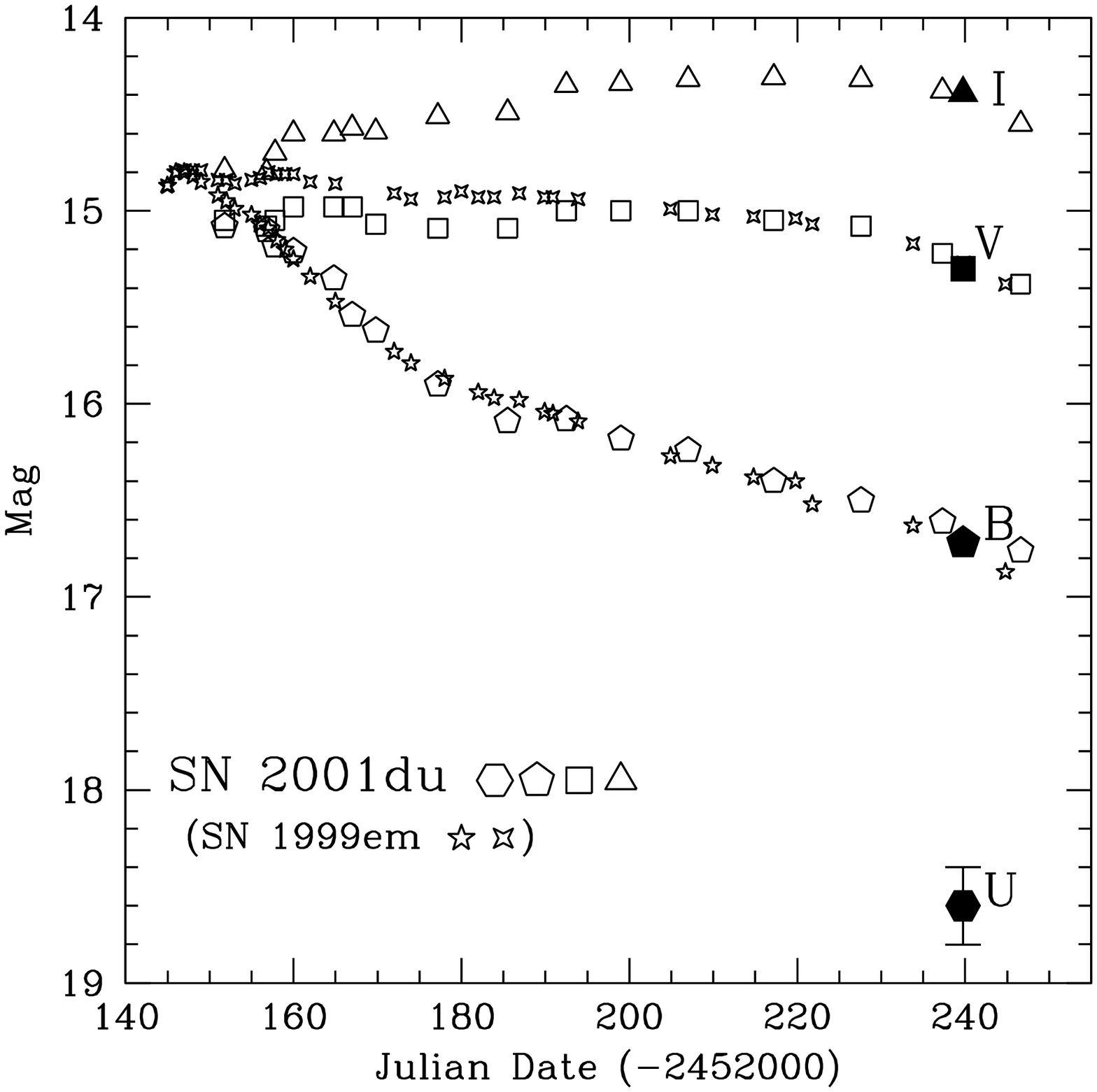}
\caption{The light curves of SN 2001du.  The {\it open pentagons\/}, {\it open
squares}, and {\it open triangles} are $BVI$ measurements from images obtained
using the YALO 1-m telescope (N. Suntzeff \& K. Krisciunas, 2003, private
communication).  The {\it filled\/} symbols are the measurements from the {\sl
HST\/} images.  Also shown for comparison are the measurements in $B$ ({\it
five-pointed stars}) and $V$ ({\it four-pointed stars}) for the Type II-P SN
1999em from Leonard et al.~(2002a), adjusted by $\sim$1 mag to match the $V$
peak for SN 2001du.}
\end{figure}

\clearpage

\begin{figure}
\figurenum{4}
\plotone{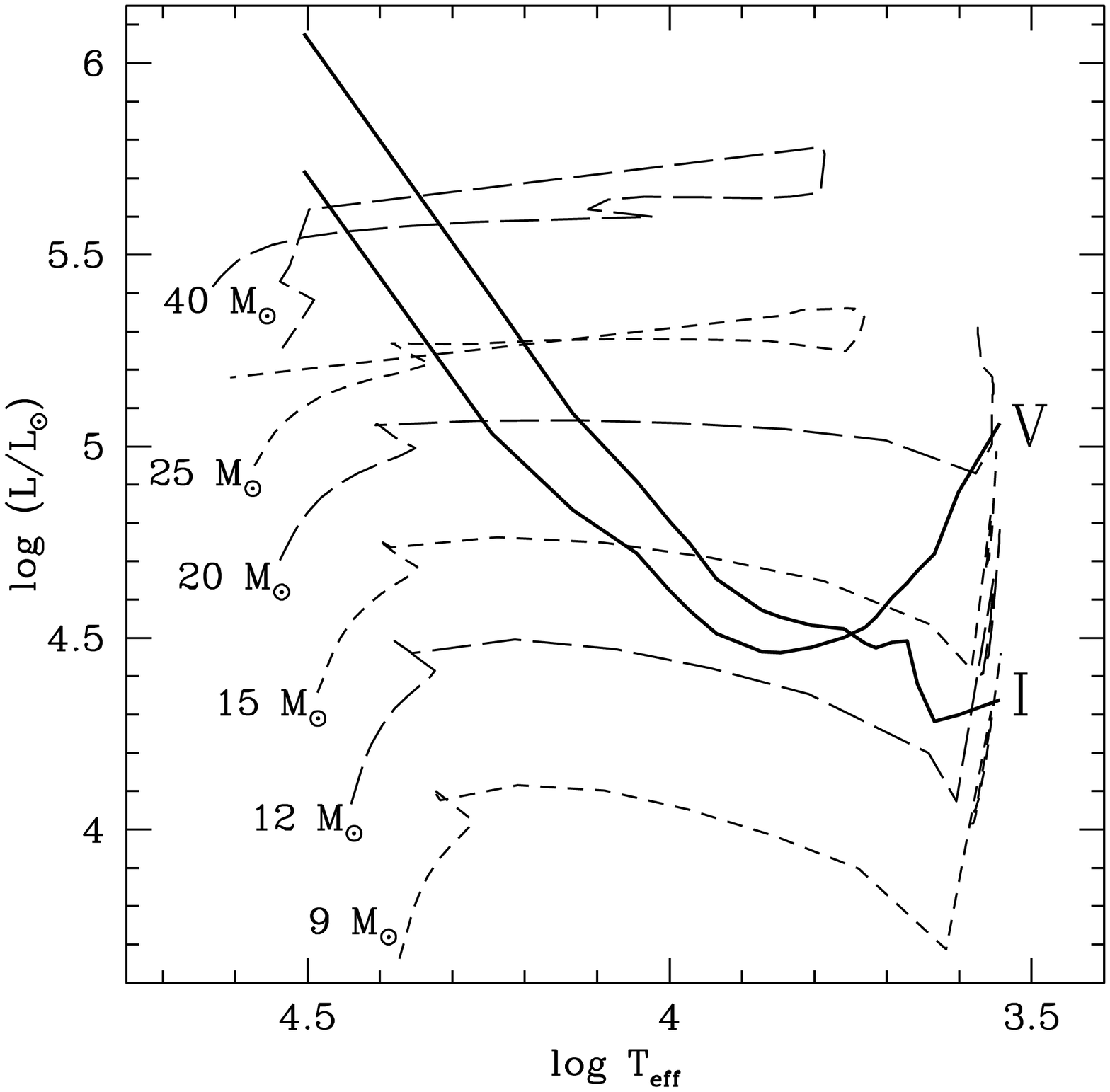}
\caption{Hertzsprung-Russell diagram showing model stellar evolutionary tracks
(alternating {\it long-dashed lines\/} and {\it short-dashed lines}) for a
range of masses from Lejeune \& Schaerer (2001), with enhanced mass loss for
the most massive stars and a metallicity $Z=0.04$.  Also shown are the
luminosity limits ({\it heavy solid lines}) for both the F555W ($V$) and F814W
($I$) bands.  Stars with luminosities brighter than these limits should have
been detected in the pre-SN {\sl HST\/} images.}
\end{figure}

\end{document}